\documentclass[%
amsmath,amssymb,
twocolumns,
reprint,%
superscriptaddress,
showpacs,
10pt,
]{revtex4-1}

\usepackage{amsmath}
\usepackage{graphicx}
\usepackage{bm}
\usepackage{amssymb}
\usepackage{amsfonts}
\usepackage{amsthm}

\begin{document}

\title[Tackling the Limits of Optical Fiber Links]{Tackling the Limits of Optical Fiber Links}

\author{Fabio Stefani}
\affiliation{ 
Laboratoire National de M\'{e}trologie et d'Essais-Syst\`{e}me de R\'{e}f\'{e}rences Temps-Espace, UMR 8630 Observatoire de Paris, CNRS, UPMC, 61 Avenue de l'Observatoire, 75014 Paris, France}%
\affiliation{
Laboratoire de Physique des Lasers, Universit\'{e} Paris 13, Sorbonne Paris Cit\'{e}, CNRS, 99 Avenue Jean-Baptiste Cl\'{e}ment, 93430 Villetaneuse, France
}%
\author{Olivier Lopez}
\affiliation{
Laboratoire de Physique des Lasers, Universit\'{e} Paris 13, Sorbonne Paris Cit\'{e}, CNRS, 99 Avenue Jean-Baptiste Cl\'{e}ment, 93430 Villetaneuse, France
}%
\author{Anthony Bercy}
\affiliation{
Laboratoire de Physique des Lasers, Universit\'{e} Paris 13, Sorbonne Paris Cit\'{e}, CNRS, 99 Avenue Jean-Baptiste Cl\'{e}ment, 93430 Villetaneuse, France
}%
\affiliation{ 
Laboratoire National de M\'{e}trologie et d'Essais-Syst\`{e}me de R\'{e}f\'{e}rences Temps-Espace, UMR 8630 Observatoire de Paris, CNRS, UPMC, 61 Avenue de l'Observatoire, 75014 Paris, France}%
\author{Won-Kyu Lee}
\affiliation{
Laboratoire National de M\'{e}trologie et d'Essais-Syst\`{e}me de R\'{e}f\'{e}rences Temps-Espace, UMR 8630 Observatoire de Paris, CNRS, UPMC, 61 Avenue de l'Observatoire, 75014 Paris, France}%
\affiliation{
Korea Research Institute of Standards and Science, Daejeon, 305-340, Korea
}%
\author{Christian Chardonnet}
\affiliation{
Laboratoire de Physique des Lasers, Universit\'{e} Paris 13, Sorbonne Paris Cit\'{e}, CNRS, 99 Avenue Jean-Baptiste Cl\'{e}ment, 93430 Villetaneuse, France
}%
\author{Giorgio Santarelli}%
\altaffiliation[Present address:]{ Laboratoire Photonique, Num\'{e}rique et Nanosciences, UMR 5298, Universit\'{e} de Bordeaux 1, Institut d'Optique and CNRS, 1, Rue F.\,Mitterrand, 33400 Talence, France}
\author{Paul-Eric Pottie}
\email{paul-eric.pottie@obspm.fr.}
\affiliation{ 
Laboratoire National de M\'{e}trologie et d'Essais-Syst\`{e}me de R\'{e}f\'{e}rences Temps-Espace, UMR 8630 Observatoire de Paris, CNRS, UPMC, 61 Avenue de l'Observatoire, 75014 Paris, France}
\author{Anne Amy-Klein}
\affiliation{
Laboratoire de Physique des Lasers, Universit\'{e} Paris 13, Sorbonne Paris Cit\'{e}, CNRS, 99 Avenue Jean-Baptiste Cl\'{e}ment, 93430 Villetaneuse, France
}%
\date{\today}

\begin{abstract}
We theoretically and experimentally investigate relevant noise processes arising in optical fiber links, which fundamentally limit their relative stability. We derive the unsuppressed delay noise for three configurations of optical links\,: two-way method, Sagnac interferometry, and actively compensated link, respectively designed for frequency comparison, rotation sensing, and frequency transfer. We also consider an alternative two-way setup allowing real-time frequency comparison and demonstrate its effectiveness on a proof-of-principle experiment with a 25-km fiber spool. For these three configurations, we analyze the  noise arising from uncommon fiber paths in the interferometric ensemble and design optimized interferometers. We demonstrate interferometers with very low temperature sensitivity of respectively  $-2.2$, $-0.03$ and $1$ fs/K. We use one of these optimized interferometers on a long haul compensated fiber link of 540\,km. We obtain a relative frequency stability of $3\cdot 10^{-20}$ after 10,000 s of integration time.
\end{abstract}

\maketitle

\section{Introduction}
Optical fiber links for frequency and time dissemination have been developed fast for the last 5 years demonstrating optical frequency transfer below $10^{ -13}$ at one second integration time over typical distance of 1\,000 km\,\cite{Newbury:2007,Fujieda:2011,Lopez:2012,Droste:2013,Marra:2012,Calosso:2014,Krehlik:2012, Calonico:2014}. The most demanding application in terms of stability and uncertainty is the direct comparison of distant optical clocks (see for instance \,\cite{Giorgetta:2013} and references herein). Their record relative frequency stability of $10^{-18}$ in less than 1\,000 s is out of reach for satellite techniques\,\cite{Fujieda:2014}.  Fiber links are now extending towards a much wider field of application with the development of national or transnational fiber networks as for instance REFIMEVE+ in France and LIFT in Italy. Frequency transfer with fiber links is expected to play a key role for fundamental physics and high resolution atomic and molecular physics\,\cite{Matveev:2013,Chanteau:2013} and for syncing large area networks as for astrophysics applications. 

Fiber links operated in the optical frequency domain use an ultra-stable laser as a frequency signal. This signal is degraded by the propagation delay fluctuations, due to thermal and mechanical perturbations that affect the optical length of the link setup. Although these perturbation are passively or actively compensated in usual fiber links configurations, a fine understanding of the phase noise contributions of a fiber link will enhance the resolution of the frequency comparison and transfer and reduce the integration time. Our aim is also to improve further the performances of fiber links for new applications, such as Sagnac interferometry on giant loops\,\cite{Schiller:2013,Clivati:2013}. 

In this paper we first describe three configurations of fiber links used for frequency comparison, rotation sensing of the inertial frame, and frequency transfer. We derive the rejection of the propagation noise in the case of a noise process with spatial correlation. Then we study the so-called interferometer noise arising from uncommon optical paths for both directions and its contribution to the overall budget. Complementary to the work reported in\,\cite{Williams:2008}, we investigate interferometric noise of fiber-based interferometers, which is related to the temperature sensitivity of the interferometers. We realize three interferometers, one for each link configurations, and demonstrate their very low temperature sensitivity.  We finally demonstrate a 540-km optical link with such an ultra-low noise interferometer and show that the interferometric noise contribution at long integration time is at $10^{-20}$ level.

\begin{figure*}
\includegraphics[width=85mm]{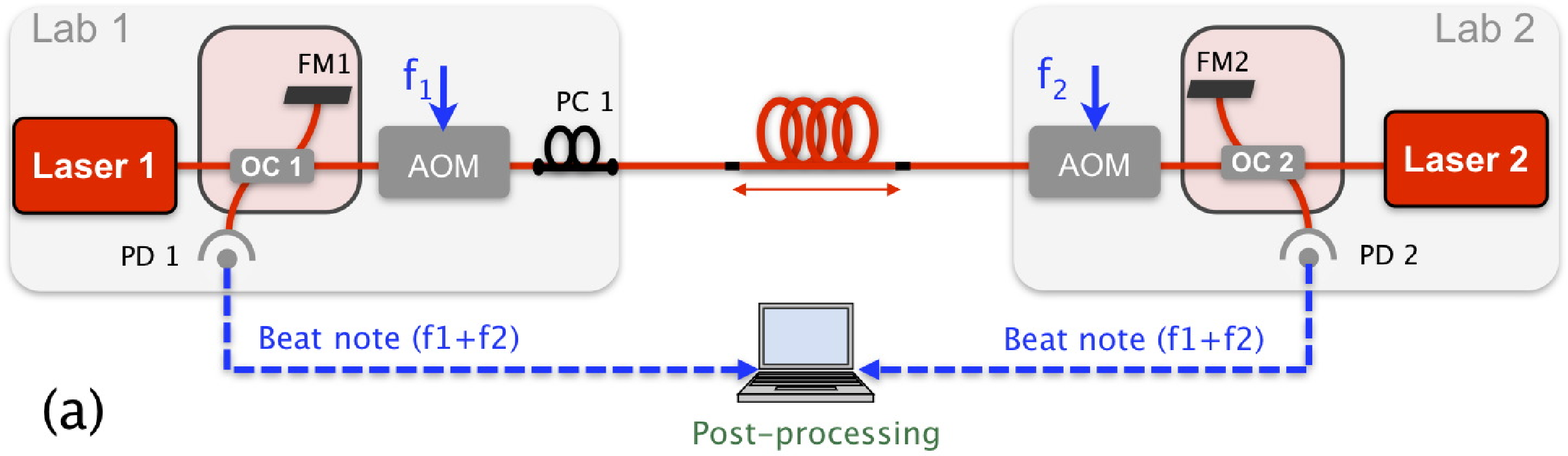} \vline \hspace{1mm}\includegraphics[width=75mm]{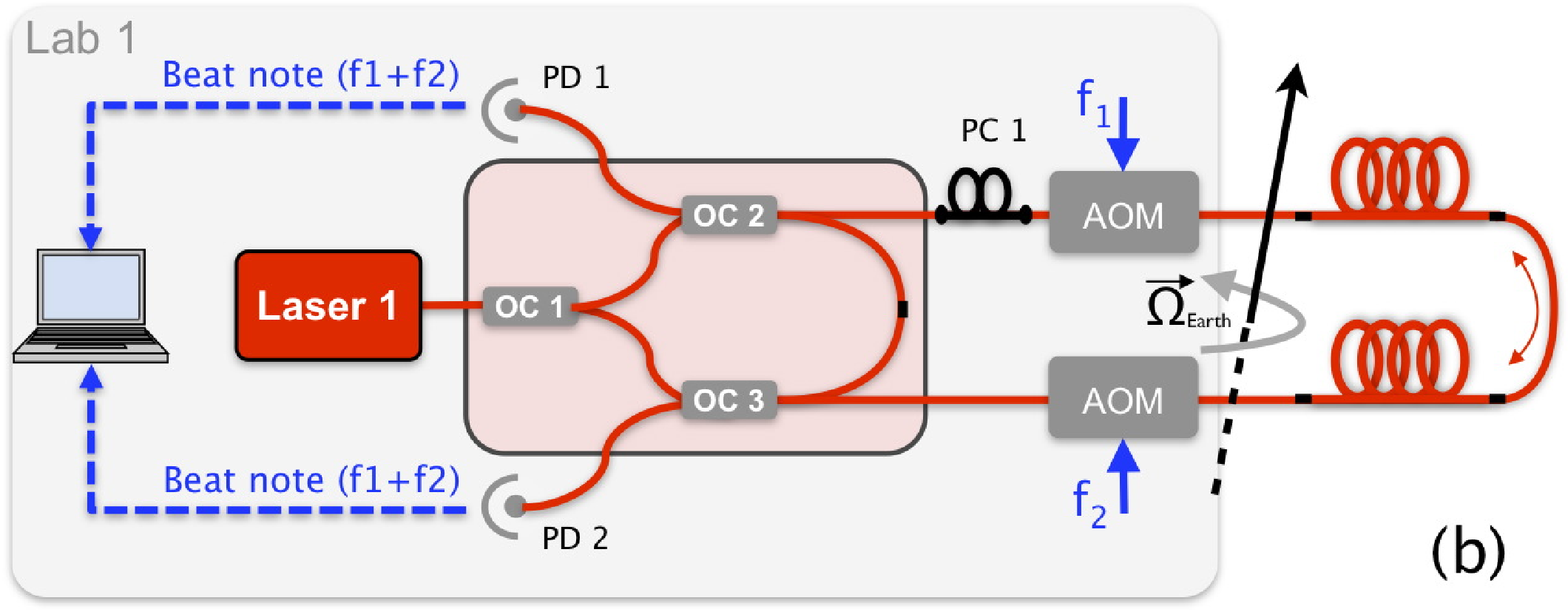}
\line(1,0){500}\vspace*{2mm}
\includegraphics[width=12cm]{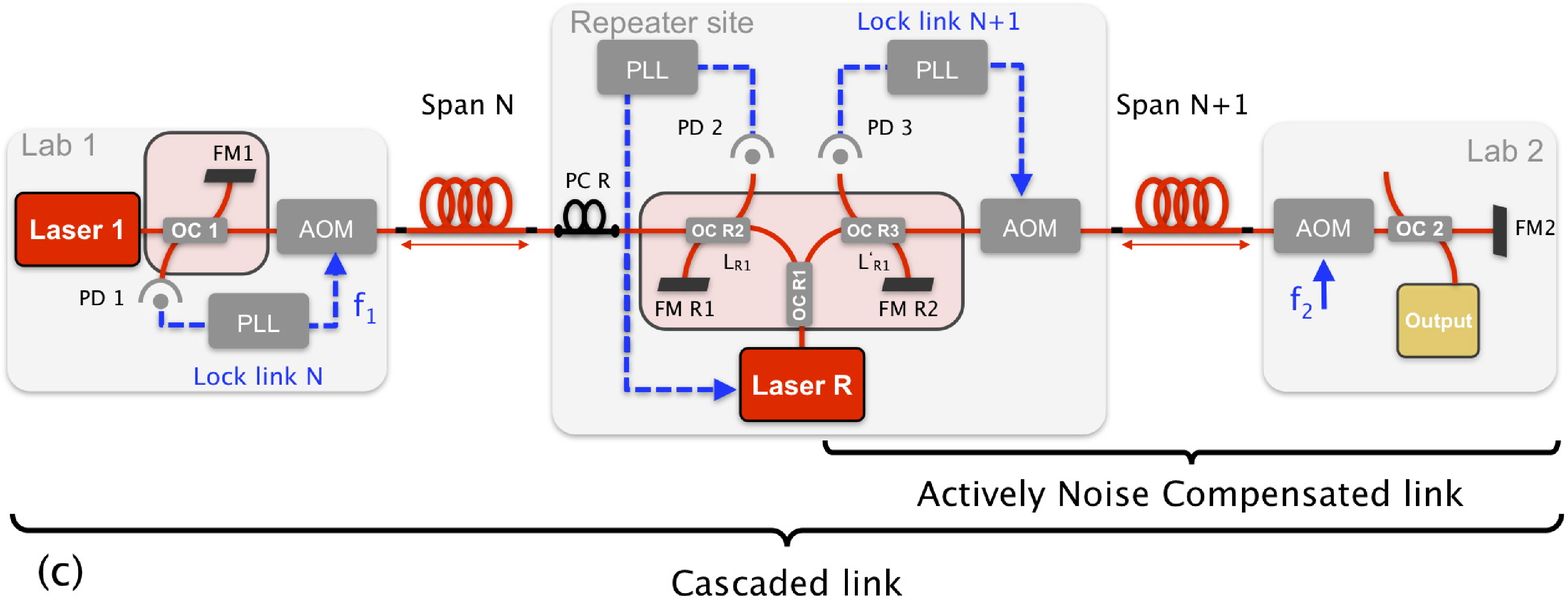}
 \caption{\label{Fig:sketches}  Experimental setups for (a) Two-way frequency comparison (b) Sagnac interferometry (c) Actively Noise Compensated link and cascaded link. FM: Faraday mirror. OC: optical coupler. AOM: acousto-optic modulator. PLL: phase-lock loop. PC: polarization controller. $\Omega_\textrm{Earth}$ is the rotation rate of the Earth.}
\end{figure*}
\section{LINK TOPOLOGY}
We consider three configurations of optical links, all of them enabling the compensation of the propagation noise. These three configurations are respectively designed for frequency comparison, rotation sensing and frequency transfer. Their setups are depicted on Fig.\,\ref{Fig:sketches}. 
For frequency comparison of remote clocks, the most simple method consists in implementing a two-way link. This method was introduced by\,\cite{Hanson:1989} and extended to optical links\,\cite{Williams:2008,Calosso:2014,Bercy:2014b}. At each end of the link, a laser is sent to the other end. One detects the frequency difference between the local laser signal and the remote laser signal by measuring their beat note signals using a Michelson-type interferometer (see Fig.\,\ref{Fig:sketches}a). At each end, the beat note interferometric detection setup is made of an optical coupler, a short fiber arm terminated by a Faraday mirror and a photodiode. The fiber link itself is the second arm of the interferometer. Assuming that the long arm's fiber noise is equal for the two directions of propagation, one can efficiently reject the delay fluctuations by synchronizing and then post processing the data, simply by dividing by two the subtraction of the data sets that were recorded at the two ends. Note that accurate frequency comparisons require an accurate synchronization of the frequency data. This requirement is about the microsecond in most practical cases, which is not that stringent, and depends on the frequency drift of the two lasers and on the length of the link\,\cite{Raupach:2012}.

When the fibre link follows a loop as depicted in Fig.\,\ref{Fig:sketches}b, starting and ending in the same lab and enclosing a non-zero effective area, an optical link can be turned into a giant Sagnac interferometer \,\cite{Sagnac:1911,Arditti:1981,Schiller:2013,Clivati:2013}. A Sagnac phase can be detected by splitting a single laser in two optical signals, injecting them clockwise and counterclockwise in the loop, and by applying the two-way compensation method  described previously\,\cite{Davis:1981}. The phase signal arises from the phase comparison between the long and the short (reference) loop. This approach was revisited recently in \,\cite{Clivati:2013}. Compared to the two-way link, the experimental setup is even simpler, because the data sampling synchronization is made trivial by the colocation of the measurements. Even more, an unstabilized laser source can be used as a reference laser, since all the laser-related phase and frequency noises are rejected in common mode.

The two-way method for frequency comparison is ineffective when the transfer of an ultra-stable frequency from a reference laboratory to a distant user is required, as when an end user has no stable clock in the laboratory. Thus, an active method should be used to compensate in real-time for the propagation noise along the fiber\,\cite{Ma:1994}. The principle of the method is sketched on the right part of Fig.\,\ref{Fig:sketches}c. The optical signal sent at one end travels forth and back through the same fiber, to allow the detection of a beat note of the local laser with the round-trip signal. AOMs are used at the two ends to distinguish the round-trip signal from parasitic back reflections. The beat note signal is a measurement of the round-trip phase noise. Assuming again that the accumulated fiber noise over the link is equal forth and back, one can then actively compensate the propagation noise by correcting the input signal phase with half of the opposite of the measured phase noise. This is done with the acousto-optic modulator at the link input end.

For all these setups the propagation noise is detected through a time delayed interferometer with strongly length-unbalanced arms. One arm consists of the fiber link, the other one is a local short reference arm. In order to improve the signal to noise ratio, an heterodyne detection of the phase difference between the two arms is implemented, by shifting the frequency in the long-arm with an AOM. The optical propagation delay in the long-haul link is a limiting factor for the noise correction. Actually both the correction bandwidth and the noise amplitude rejection are limited by the round-trip propagation time \,\cite{Williams:2008}. For instance a link of 1840\,km leaves the acoustic noise almost uncorrected as the bandwidth is below 20\,Hz\,\cite{Droste:2013}. In order to circumvent this issue and to face the large losses of public network, our group introduced cascaded link, where the propagation noise is compensated by successive and consecutive fiber spans\,\cite{Jiang:2008,Lopez:2010}. With this approach, the optical signal has to be repeated from one span to the other using a so-called repeater station. In this setup, a local laser is phase-locked to the incoming signal of the previous span in order to clean up and amplify the metrological signal, as depicted on Fig.\,\ref{Fig:sketches}c. 

Part of the fiber noise cannot be rejected by the noise compensation system. This noise arises from the interferometric detection of the fiber noise\,\cite{Williams:2008,Grosche:2009}. It can be due for instance to propagation noise in the reference arm. The main noise sources of fiber-based interferometers are thermal effect. Addressing this noise is crucial for the design of the optical part of a station, in which the laser phase should be well defined and should not suffer any thermal fluctuations, as explained below. 

Apart from this interferometric noise and unsuppressed delay noise that will be addressed successively in the next two parts, the fiber noise that is accumulated from the output of the optical oscillators to the input of the interferometric setup is also contributing to the noise budget. It is indistinguishable from the intrinsic noise of the optical oscillators itself. It remains even if the unsuppressed delay noise and the interferometric noise are zero. To get rid of that, one should consider very short fiber paths from the oscillator to the interferometer, very good thermal stabilization and low acoustical noise levels, or a short compensated link. This noise can be neglected in a case of a demonstration setup for frequency comparison or dissemination (where an optical oscillator is compared to itself), and in any case for a giant Sagnac interferometer. It will not be considered below. 

\section{UNSUPPRESSED DELAY NOISE}
\label{sec:unsuppressed delay noise}
\label{subsec:cln}

\subsection{Free fiber noise}
Following Newbury and coworkers, we will consider a phase noise density per unit of length $\delta \varphi(z,t)$, that depends of time $t$ and position $z$ along the link\,\cite{Newbury:2007}. $\delta \varphi(z,t)\,dz$ gives the phase noise of the fiber elementary length $dz$. Thus the fiber noise at time $t$ arising from the propagation from the end 1 at position $z=0$ to the end 2 at position $z=L$ is given by $\phi_{\textrm{12}}(t) =\int_0^{L} \delta \varphi(z, t-T+z/v)\,dz$, where $v$ is the light velocity in the fiber and $T =L/v$ the one-way propagation delay.

Considering standard random signal processing, the fiber phase noise Power Spectral Density (PSD) is given by the Fourier transform of the autocorrelation function of its phase noise. This approach allows to take into account the statistical properties of the phase noise, as it depends on the model established for the random signal $\delta \varphi(z,t)$. The full calculations are detailed in Appendix\,\ref{AppendixA}. 

Since it originates from thermal and acoustical fluctuations, the phase noise density $\delta \varphi(z,t)$ has a very short spatial correlation along the fiber. Here we assume that its spatial autocorrelation function has a narrow gaussian profile of deviation $\Delta L$, with $\Delta L\ll L$. $\delta \varphi(z,t)$ is thus practically uncorrelated for two different positions z and z' separated by more than a few $\Delta L$. It extends the previous studies in which the spatial autocorrelation of the noise was considered as an impulse function \cite{Williams:2008,Bercy:2014a}. This  calculation method can be extended to arbitrary correlation functions, in time or in space. 

For sake of simplicity, we also consider here that the fiber noise per unit length $\delta \varphi(z,t)$ has constant statistical properties over $z$. We obtain the following PSD for the free fiber phase noise $\phi_{\textrm{12}}(t)$
\begin{equation}
\label{eq:freefibernoise}
\begin{split}
S_{\textrm{12}}(f)= \Delta L L \sqrt{2\pi}e^{-\frac{( \sqrt{2}\pi f \Delta L)^2}{v^2} } S_{\delta\varphi}(f)\\
\end{split}
\end{equation}
where $S_{\delta \varphi}(f)$ is the PSD of the fiber noise per unit of length $\delta \varphi(z,t)$.

\subsection{Two-way link for frequency comparison and Sagnac interferometry for rotation sensing}\label{Sec:TW}
In the case of a two-way frequency comparison, we consider the one way fiber noise from the end 1 to the end 2, $\phi_{12}(t)$, and from the end 2 to the end 1,  $\phi_{21}(t)$. The laser phases $\phi_1(t)$ of the Laser 1 and $\phi_2(t)$ of the Laser 2 are defined as the lasers enter the interferometric setup. From the frequency measurements made on the photodiodes PD1 and PD2, one obtains in post-processing the two-way phase signals $(-\phi_\textrm{PD1} +\phi_\textrm{PD2})/2 = (\phi_1-\phi_2)+(\phi_{12}-\phi_{21})/2$, where the time dependency has been omitted for sake of simplicity. The ideal situation for a frequency comparison is when $\phi_1-\phi_2$ is the dominant term. The performance of the comparison is asset when the two lasers are perfectly coherent, {\it i.e.} when $\phi_1=\phi_2$. For this study focused on the fiber noise contribution we conveniently defined the two-way phase signal as $\phi_{\textrm{tw}}(t)=(\phi_{12}(t)-\phi_{21}(t))/2$. Considering that $2 \pi f T \ll 1$, and for a first order expansion, one has\,: 
\begin{equation}\label{Eq:phiTW}
 \phi_{\textrm{tw}}(t)=\int_0^L{\left(\frac{z}{v}-\frac{T}{2}\right)} \delta \varphi ' (z,t)dz
\end{equation}
One deduces that the two-way phase PSD is  then\,: 
\begin{equation}
\label{eq:twoway}
\begin{split}
S_{\textrm{tw}}(f)=\frac{1}{12} (2 \pi f T)^2 \cdot e^{+\frac{(\sqrt{2} \pi f \Delta L)^2}{v^2} } \cdot S_{\textrm{12}}(f) 
\end{split}
\end{equation}
The full calculation is detailed in Appendix\,\ref{AppendixA}. 

It is convenient to limit our observations to a maximum frequency $f_H$ for which $\frac{1}{12}(2 \pi f_H T)^2=1$, in order to avoid that the post-processed two-way noise overtakes the original one way noise. In that case, it can be easily demonstrated that the exponential term in Eq.\,\ref{eq:twoway} is bounded to $e^{+{6 \cdot (\Delta L / L)}^2}$, making it negligible for every practical realization. 

The same formulas are found for a Sagnac configuration. 

\subsection{Actively noise compensated link for frequency dissemination}
In the case of an actively noise compensated (ANC) link, and for a first order expansion, the compensated fiber noise at output end 2 is $\phi_{\textrm{anc}}(t) =\int_0^{L} z/v \delta \varphi '(z, t)\,dz$\,\cite{Williams:2008,Bercy:2014a}.
One can demonstrate that the PSD of  $\phi_{\textrm{anc}}(t)$ is\,:

\begin{equation}\label{Eq:SphiCompToSphiLibre}
S_{\textrm{anc}}(f)=\frac{1}{3}(2 \pi f T)^2 \cdot e^{+(\frac{\sqrt{2} \pi f \Delta L}{v} )^2} \cdot S_{\textrm{12}}(f) 
\end{equation}
The calculation is detailed in Appendix\,\ref{AppendixA}. That's almost the same formula that the one derives in\,\cite{Newbury:2007}, with an additional exponential c\oe fficient arising from the spatial correlation of the noise. According to the theory of servo-loops, the active noise compensation system can operate up to a frequency approximatively equal to $1/(4T)$, setting also an upper limit for the validity of Eq.\,\ref{Eq:SphiCompToSphiLibre}\,\cite{Williams:2008}. Beyond this limit, the fiber noise remains uncorrected. Considering this, we can derive that the exponential coefficient in Eq.\,\ref{Eq:SphiCompToSphiLibre} can reach a maximum value of $e^{ (\pi^{2}/8)  (\Delta L / L)^2}$, which is negligible in practice to the best of our knowledge.

From Eq.\ref{Eq:SphiCompToSphiLibre} and following considerations, one expects that the amplitude and the bandwidth of the noise correction are decreasing linearly with $L$. The performance of the link can then be improved by dividing a link into smaller subsections. This is the cascaded link approach\,\cite{Jiang:2008,Lopez:2010}. For instance, dividing a link in $N$ independent subsections of equal length, and considering that the power of the free fiber noise scales as the length of each subsection, both the amplitude and the bandwidth of the noise rejection are increased by a factor $N$.

\subsection{Two-way link with noise compensation}
A question that arises therefore is whether a composite system can improve further the fiber noise rejection. Several configurations are considered here. We first consider the case of an active compensation system set at one end of a two-way setup. We demonstrate in Appendix\,\ref{AppendixB} that it does not lead to a higher noise rejection. On the contrary, one finds that the two-way phase noise PSD is indeed 4 times higher than that of a post-processed two-way without active  compensation. We also consider the case of a double round-trip noise detection system at each end, and a post-processing correction of each end phase signal. We demonstrate in Appendix\,\ref{AppendixB} that the two post-corrections at each ends canceled each other and that this configuration does not lead to higher noise rejection. 

\subsection{ Real-time two-way link}\label{Sec:TWRT}
\begin{figure}
 \includegraphics[width=8cm]{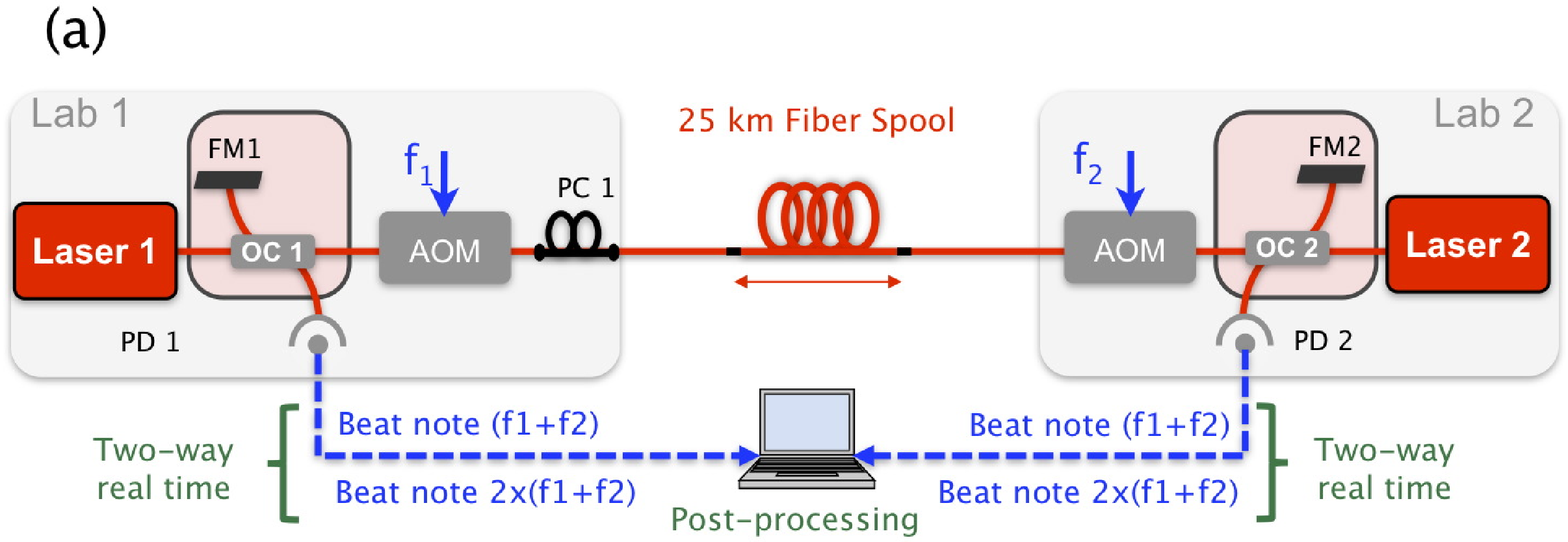}
 \includegraphics[width=8cm]{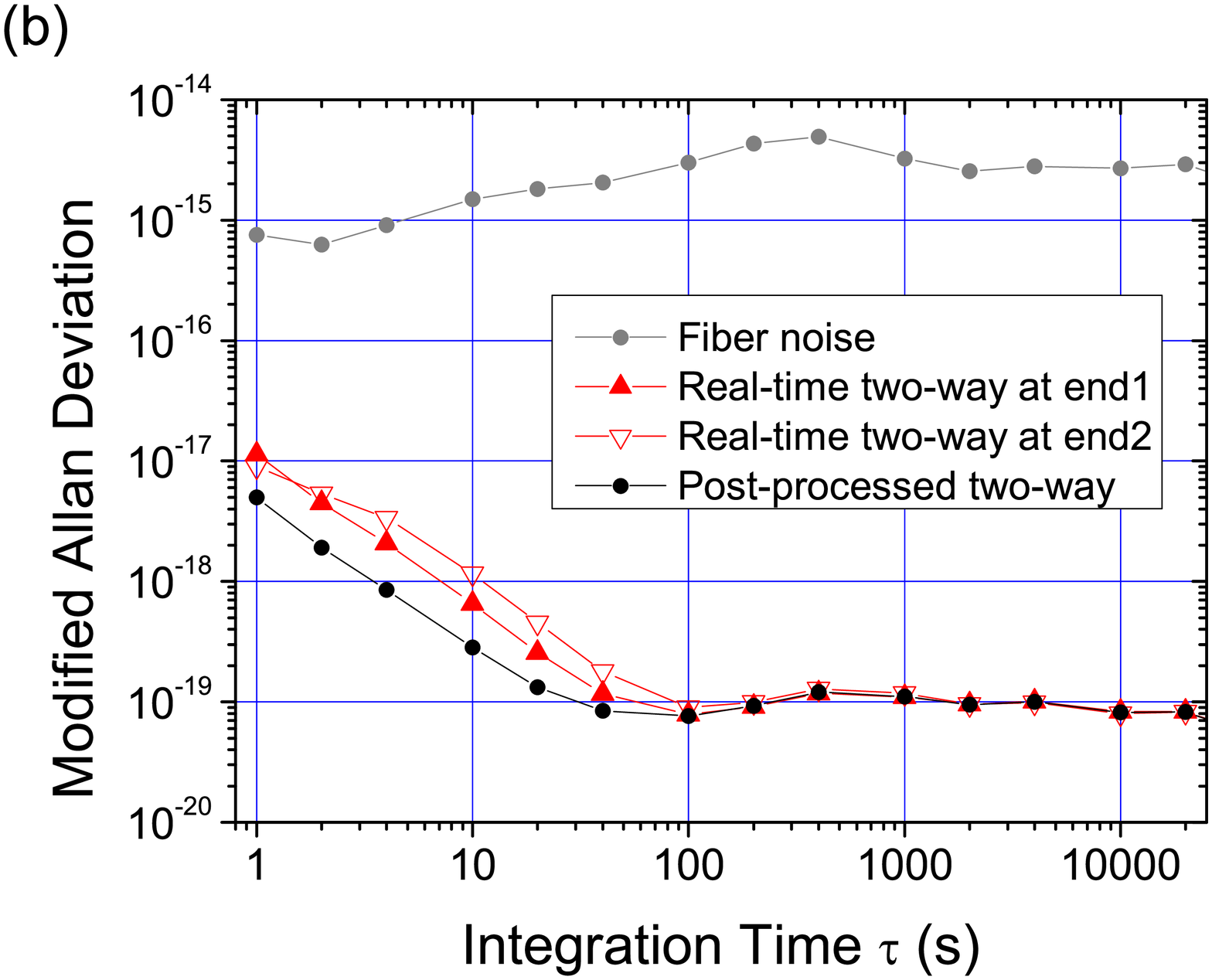}
 \caption{\label{Fig:sketch2wayRT} (a) Experimental sketch of the real-time two-way setup. FM: Faraday mirror. OC: optical coupler. AOM: acousto-optic modulator. PC: polarization controller. (b) Modified Allan Deviation of the recorded data in real-time on PD1 and PD2 (red filled-up triangles and red open down-trainagles), in post-processing from signals on PD1 and PD2 (black circles), and free-running fiber noise (gray circles).}
\end{figure}

In this subsection we consider an alternative two-way setup which takes advantage of the round-trip noise detection performed with the ANC configuration. This technique was introduced in\,\cite{Bercy:2014b} for a two-fiber setup where the light travels the fibers with uni-directional propagation. We introduce here a bi-directional configuration with a single fiber. If one flips at both ends the play role of the Faraday mirrors and the photodiodes of a two-way setup (see Fig.\,\ref{Fig:sketches}a and Fig.\,\ref{Fig:sketch2wayRT}a), one has the possibility to record simultaneously on each photodiode two signals, that we labelled {\it A} and {\it B}.  One can write the beat notes detected by the photodiodes PD1 and PD2 as follows\,:
\begin{eqnarray}
   \left\{
    \begin{aligned}
 	\phi_{\textrm{PD1}_A}  &=(\phi_2+\phi_{21})-\phi_1 &\textrm{at frequency }F \\
 	\phi_{\textrm{PD1}_B} &=(\phi_{12}+\phi_{21}) &\textrm{at frequency }2F  \\
 	\phi_{\textrm{PD2}_A} &=(\phi_1+\phi_{12})-\phi_2 &\textrm{at frequency }F \\
 	\phi_{\textrm{PD2}_B} &=(\phi_{12}+\phi_{21}) &\textrm{at frequency }2F \\
    \end{aligned}
    \right.
 \label{Eq:set2wayBiDirRT}
\end{eqnarray}
where $\phi_1$, $\phi_2$, $\phi_{21}$, and $\phi_{12}$ are defined in section\,\ref{Sec:TW}. We set $F=f_1+f_2$. The signals labelled as {\it A} are the beat note signals of the local laser with the remote laser that propagated in single trip over the fiber. Its mean value is the sum of the AOM's frequencies, $F$. The signals labelled as {\it B}  are the beat notes of the local laser with itself after a round trip. Its mean frequency is twice the sum of the AOM's frequencies, $2F$. The signals labelled as {\it A} in Eq.\,\ref{Eq:set2wayBiDirRT} from the two photodiodes at each end can be post-processed, and gives the usual two-way phase signal\,:
\begin{equation}\label{Eq:2way-b}
(-\phi_{\textrm{PD1}_A} +\phi_{\textrm{PD2}_A})/2 = (\phi_1-\phi_2)+(\phi_{12}-\phi_{21})/2
\end{equation}

Assuming again that $\phi_{12} =  \phi_{21}$,  one finds the regular two-way noise rejection, as the fiber noise cancels out and only the laser's phase difference remains. Note that the power of the optical beat note signals in post processing is the same as for the setup of Fig.\,\ref{Fig:sketches}a. 

In addition, the same results can be obtained by combining the $2F$ and $F$ beat notes signals detected on the same photodiode. For instance at the end\,1 on PD\,1 one has\,:
\begin{equation}\label{Eq:2way-u-local}
 -\phi_{\textrm{PD1}_A} +\phi_{\textrm{PD1}_B} /2 = (\phi_1-\phi_2)+(\phi_{12}-\phi_{21})/2
\end{equation}
the same phase difference can be obtained at end\,2 on PD\,2. This approach, that we called "local two-way", allows to process in real-time the noise rejection in  both distant laboratories, relying only on data acquired locally. It avoids the necessity to exchange and synchronize data between distant sites to compute the stability of the link and diagnose any malfunction. The real-time processing can be done by a simple tracking oscillator and do not need high-speed digital electronics.

The main difference between the real-time signal and the post-processed one is that the limit set by the propagation delay is twice for the real-time data, due to the round trip. Following the same approach as for the previous setups, we demonstrate that the noise rejection for the real-time two-way is given by Eq.\,\ref{Eq:SphiCompToSphiLibre}, as for the actively compensated noise link. The post-processed two-way obeys Eq.\ref{eq:twoway}, and the noise rejection is expected to be four times higher. Details are given in Appendix\,\ref{AppendixB}. One expects then a factor of $2$ between the relative frequency stability in real-time and in post processing.

In order to check the validity of the approach, we have experimented this scheme on a fiber spool of 25\,km, with a single laser split in two with an optical fiber coupler. We used on-the-shelf fiber components with approximately matched lengths. The interferometric setup was set in a box with an insulating foam. The signals of each photodiodes are separated with RF filters, and then amplified, and filtered. The signal's frequencies are recorded with a dead-time free frequency counter, operated in $\Lambda$-mode with 1-s gate time\,\cite{KplusK}. Less than 60 points are removed from the two-way data set in post processing and from the real-time data set recorded on PD1. The second real-time data set recorded on PD2 suffers from experimental imperfections (as insufficient filtering) and has a poorer signal-to-noise ratio. 120 points are removed from the data set. The modified Allan deviation (MDEV) of the relative frequency fluctuations of the beat notes are displayed on Fig.\,\ref{Fig:sketch2wayRT}. We found a nice agreement to theory, with as expected, a factor of two between real-time two-way and the post-processed two-way. As the proof of principle is obtained with a 25-km fiber spool, the MDEV are as low as $1\cdot10^{-17}$ and $5\cdot 10^{-18}$ at 1-s  integration time for the real-time data and post-processed data respectively. The measurements reaches the interferometric noise floor between 50 and 100\,s of integration time, and the three curves perfectly merge themselves at longer integration time. This limitation can be overcome with a proper design and a careful realization of the interferometric setup, as it is discussed in the next section.

\section{INTERFEROMETRIC NOISE\label{sec:interferometric noise}} 

The noise compensation system relies on the assumption that $\phi_{12}=\phi_{21}$. This assumption can be violated by non-reciprocal noise, non reciprocal path, and non reciprocal effect (as Sagnac effect). 

In practical implementation, the measurement  of these quantities ($\phi_{12}$ and $\phi_{21}$) contains noise sources inherent to the detection system and that cannot be rejected by constitution. Actually the main noise source may arise from the instability of the interferometric setup. The case of free-space optics interferometer was addressed in\,\cite{Williams:2008}. We will focus here on fiber-based interferometric setup. 

The detected phase noise is obviously sensitive to the propagation noise in the reference arms\,\cite{Grosche:2009}. The interferometric noise can be asset by carefully writing the propagation's equations through the interferometric setup. It will be demonstrated below that the interferometric noise arises from the uncommon paths and their sensitivity to external perturbations. Acoustic and mechanical noises can be made negligible thanks to the very short lengths of the reference arms and a well designed isolation box. The temperature is therefore homogeneous inside the interferometer enclosure. Thus the so-called interferometric noise is mainly due to long-term thermal effects. 

\subsection{Fiber thermal sensitivity and first-order thermal noise coefficient}
\begin{figure}
 \includegraphics[width=8cm]{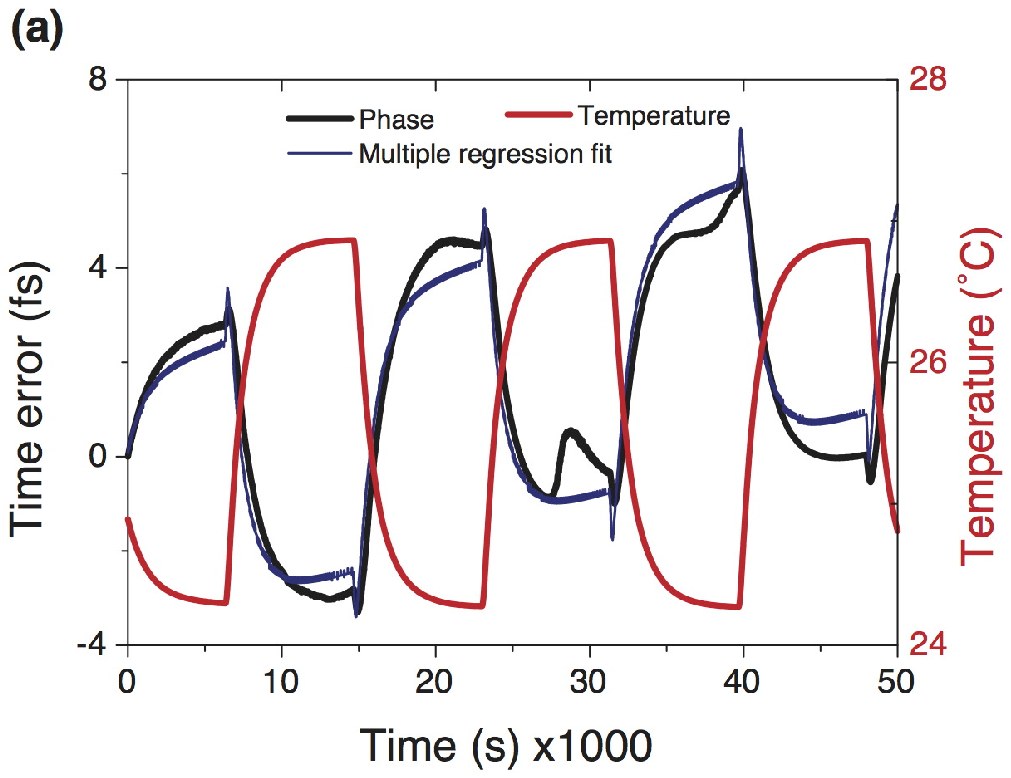}
 \includegraphics[width=8cm]{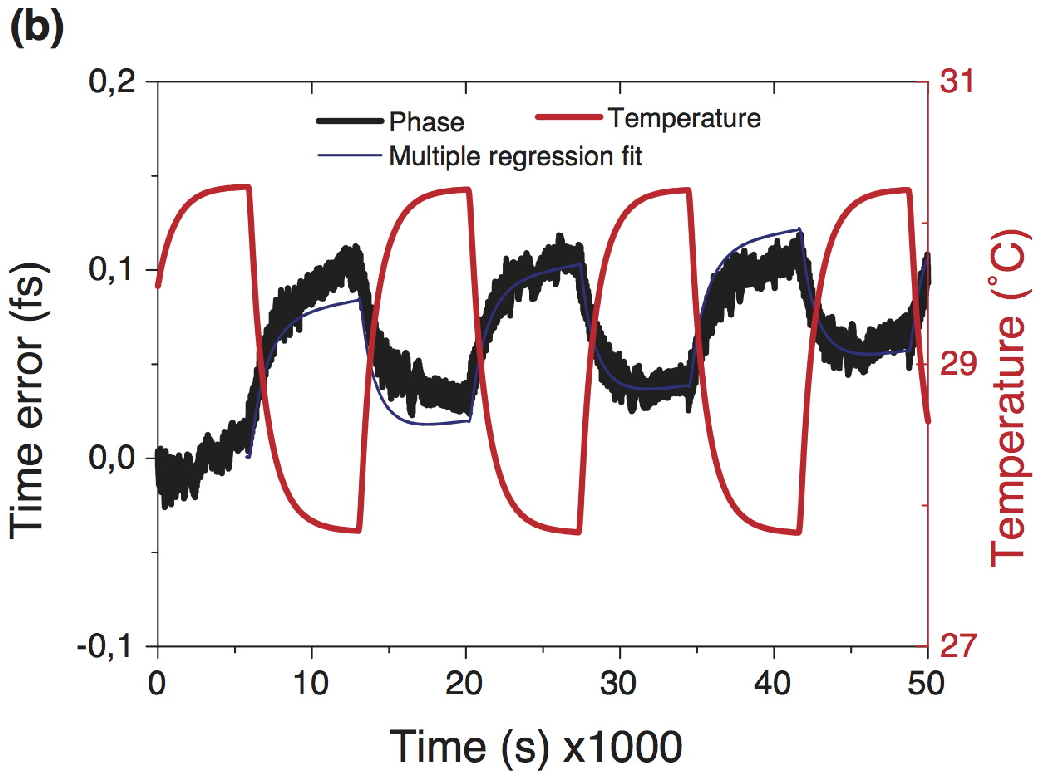}
 \includegraphics[width=8cm]{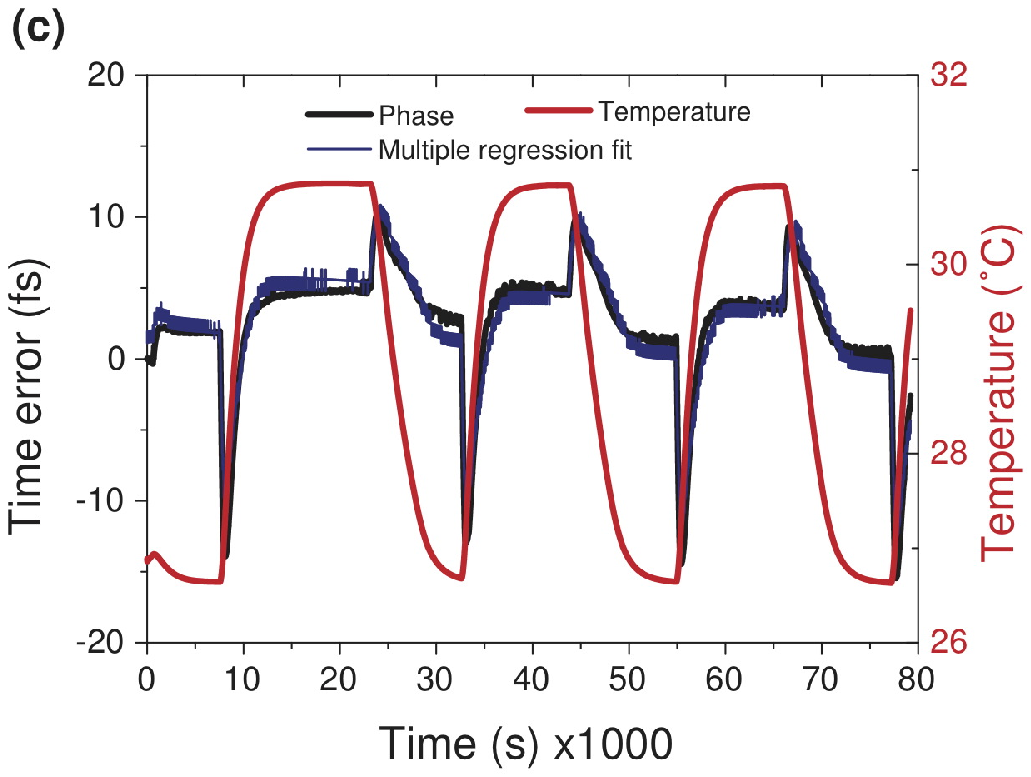}
 \caption{\label{Fig:TempSensiInterfero} Temperature modulation of the interferometer and the induced phase error at the output of the interferometer expressed in fs as a function of time. (a) Two-way interferometer (b) Sagnac interferometer (c) Cascaded link. }
\end{figure}

We estimate the long term thermal effects on the fiber phase noise density. Temperature variation acts both on the fiber physical length and on the fiber index. Around the working temperature $T_0$, the thermal fibre noise density can be expressed as a deterministic function of the temperature variation\,\cite{Cohen:1979}\,:
\begin{equation}
\label{eq:thermal phase noise}
\delta \varphi(z,t)=\omega\frac{n(T_0)}{c}\left( \alpha_{\textrm{SiO}_{2}}+\frac{1}{n(T_0)} \frac{\partial n}{\partial T}(T_0)\right)\cdot \Delta T(z,t) 
\end{equation}
where $n(T_0)=1.468$ is the fibre refraction index, $\alpha_{\textrm{SiO}_{2}}=0.55 \cdot 10^{-6} K^{-1}$ is the fiber linear expansion coefficient, $c$ the speed of light, and $\Delta T$ the variation from the reference temperature $T_0$. $\omega$ is the optical carrier angular frequency. The second term inside the parenthesis is the so-called thermo-optical coefficient and is equal to $7\cdot 10^{-6}K^{-1}$ \ \cite{chen:2008}. 
 
In order to be independent of the angular carrier frequency, we conveniently expressed the phase error as a time error in fs. Eq. \ref{eq:thermal phase noise} can be re-written as $ {\widetilde{\delta \varphi}} (z,t)=\gamma\cdot\Delta T(z,t)$, were $\gamma$ is a phase-temperature coefficient at first order. For the SMF28 fiber specifications, we obtain $\gamma= 37\,fs/(\textrm{K}\cdot\textrm{m})$ for an optical carrier at $194.4$\,THz and at $T_{0}=298$\,K.

Note that these thermal variations which affect the interferometers are quasi-static compared to the light round-trip propagation delay in a long haul fiber link. Thus the total thermal phase variation along a fiber section of length $L$ can be simply expressed as\,:
\begin{equation}
\Delta\phi(t)=\gamma \omega \int_{\textrm{0}}^{\textrm{L}}\Delta T(z,t)\,\textrm{d}z
\end{equation}

When one considers a periodic modulation affecting the phase measurements with a period $T_m$, and using the $\gamma$ c{\oe}fficient introduced above, one can show that the Allan deviation of the relative frequency modulation is as high as $1.7\cdot10^{-13}$s/(K$\cdot$m$\cdot T_m$). For a typical perturbation due to an air conditioning system, with $T_m=1\,000$\,s, 0.1 K of peak-to peak modulation depth, and 10\,cm of optical length mismatch, one expects a bump of the Allan deviation as high as $10^{-18}$ at approximately 270\,s.

In order to minimize the interferometric noise, the actual strategy is to adapt as precisely as possible the relevant optical lengths, and keep those arms at the same controlled temperature as well as possible\,\cite{Bercy:2014a,Grosche:2014}. Such a passive compensation is quite difficult to implement since the contributions of all the optical components, as for instance the Faraday mirrors, optical isolators, couplers, and AOMs, have to be taken into account. Our approach is then to measure the net phase sensitivity of the constituted interferometer. 

\subsection{Experimental setup}
In order to determine the temperature sensitivity of the interferometric setups depicted in Fig.\,\ref{Fig:sketches}, we have built a large aluminum box. The box is  isolated from the experimental breadboard with insulating foams, and shielded with a Mylar insulating film. The temperature of the interferometer is sensed by one thermistor, set in thermal contact with the interferometer. The temperature is recorded by a data logger Agilent 34970A. The box is heated by a Peltier element in thermal contact with one wall of the box. The temperature of the hot face of the Peltier element is sensed by a second thermal sensor. We applied a square modulation to the set-point temperature of the temperature stabilization loop with a period of about 15\,000\,s.

\subsection{Two-way link for frequency comparison}

We are considering first a two-way setup as depicted on Fig.\,\ref{Fig:sketches}a, designed for an optical frequency comparison. The uncommon paths are constituted by the two short arms of the two interferometers at each end. Thus the two-way phase error $\Delta\phi$ is\,: 
\begin{equation}
\Delta\phi(t)=2\gamma\omega\left(\int_{\textrm{C1}}^{\textrm{FM1}}\Delta T(z,t)\,\textrm{d}z-\int_{\textrm{C2}}^{\textrm{FM2}}\Delta T(z,t)\,\textrm{d}z\\\right)
\end{equation}
For each end of the link, the critical optical length is thus given by twice the length of the short reference arm. For an homogeneous temperature inside the interferometer, the response of the interferometer to a temperature step will be minimized when these optical lengths are minimized and made equal at each end. Experimentally, we have measured the temperature sensitivity of such interferometer, by placing two independent interferometers on the same heating device, exciting the interferometer with temperature step of 4\,K. The phase error is simultaneously measured and converted to time error and expressed in fs. We plot the results of the experiment in Fig. \ref{Fig:TempSensiInterfero}a. Neglecting one glitch that occurs around 30\,000 s, we found from a multiple linear regression fit a temperature sensitivity of -2.2\,fs/K, a sensitivity to the temperature derivative of $+0.8$ps/(K/s), and a time error drift of $+1\cdot 10^{-4}$\,fs/s. 
This temperature sensitivity is the best value reported so far for a two-way link with such an interferometer, to the best of our knowledge. 

This two-way interferometers were tested on a 100-km fiber loop surrounding Paris. The full experimental description and the frequency comparisons performances are reported in\,\cite{Bercy:2014b}. A flicker noise floor as low as $5.10^{-21}$ at 30\,000 s integration time was reached with this interferometric setup. 

We note also that a real-time two-way interferometer as introduced in Section\,\ref{Sec:TWRT} may have exactly the same performances as this one, as the difference consists only in the respective positions of the Faraday mirrors and the photodiodes.

\subsection{Two-way closed loop link for Sagnac measurement}
Sagnac interferometers are that special they are made to measure phase shift in one laboratory, as sketched in Fig.\,\ref{Fig:sketches}b. By contrast with a two-way setup, there is only one interferometer and two beat-notes. Thus there is no uncommon fiber paths. 

We realized experimentally such an interferometer, and we have measured its temperature sensitivity with the same experimental apparatus. We excited the interferometer with temperature step of 2\,K.  We plot the results of the experiment Fig. \ref{Fig:TempSensiInterfero}b. We apply again a multiple regression fit to the data, and we found a very low temperature sensitivity of -0.03\,fs/K, a sensitivity to the temperature derivative of $+1.4$\,fs/(K/s), and a time error drift of $+1.3\cdot 10^{-6}$\,fs/s, as expected from this kind of interferometers.
 
\subsection{Compensated link for frequency dissemination}\
\begin{figure}
 \includegraphics[width=8cm]{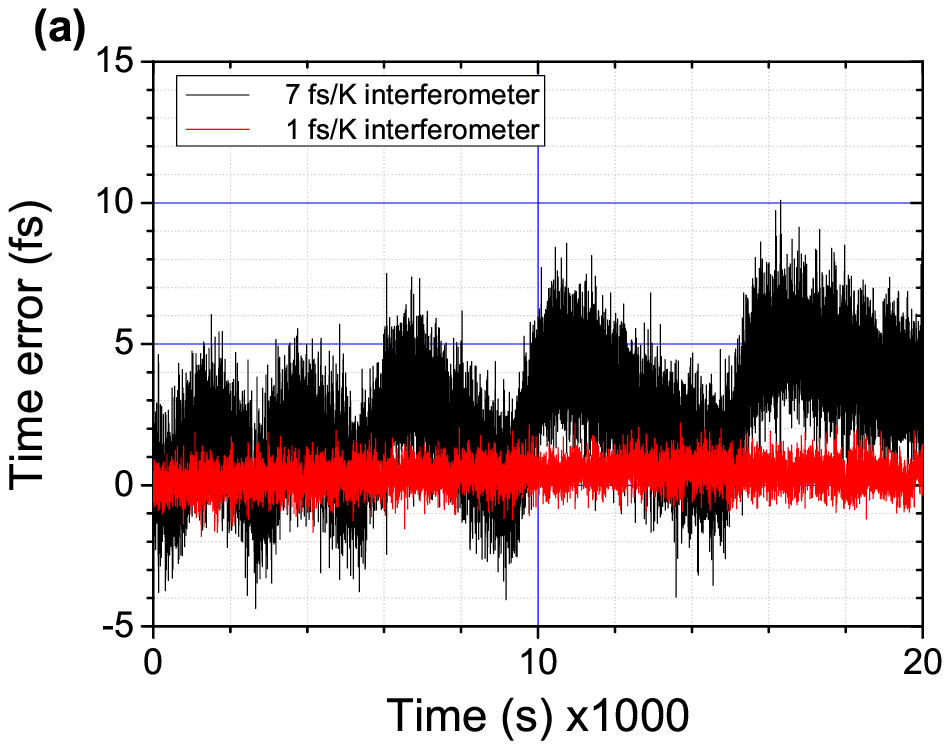}
 \includegraphics[width=8cm]{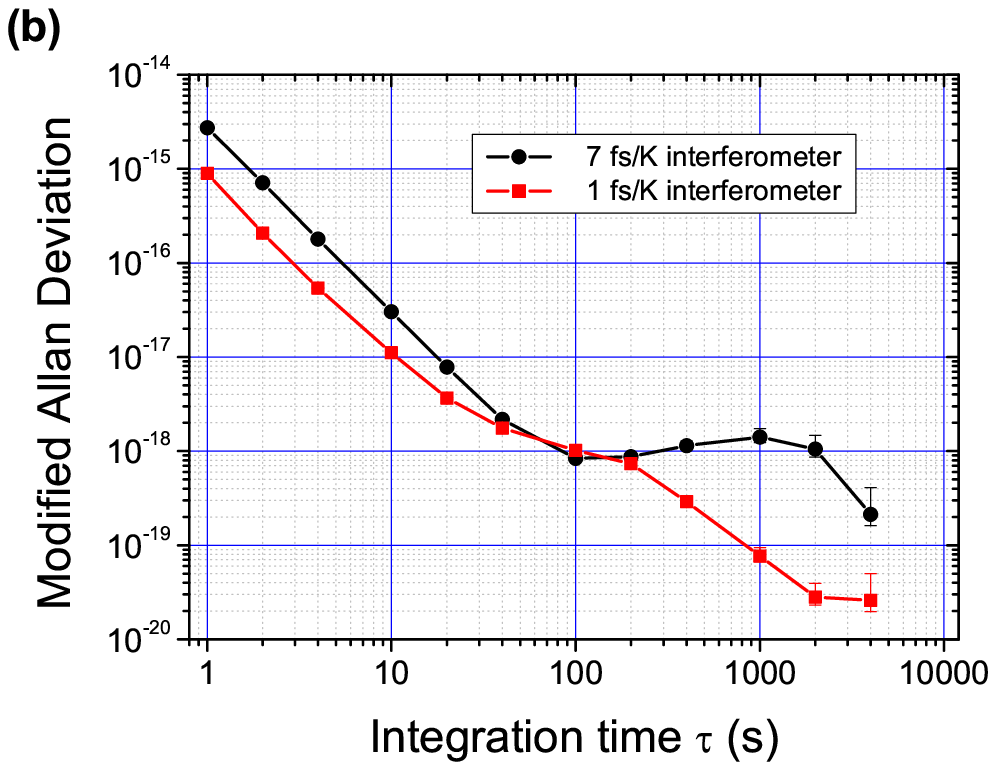}
 \caption{\label{Fig:fig3-results540km}\label{Fig:fig4-MDEV} Comparison of the transferred phase with an ANC setup with two interferometers, with a sensitivity of 7 fs/K (black trace) and 1 fs/K (red trace). (a) Time error data as a function of the measurement time. (b) Modified Allan deviation as a function of the integration time.}
\end{figure}

We are considering the interferometer of a repeater laser station of a cascaded link\,\cite{Lopez:2012}. The experimental setup is depicted on Fig.\,\ref{Fig:sketches}c. Detailed descriptions of the experimental setup are given in\,\cite{Lopez:2010,Lopez:2012}. For sake of clarity, we depicted only the main optical elements. The principle of operation is to successively phase lock a repeater laser (RL) to the upcoming signal arriving from an ultra-stable laser after propagation on the span $N$, using their beat note detected on PD2. The repeater laser copy the phase of the ultra-stable laser, when the span $N$ is actively stabilized and within the bandwidth of the loops. The repeater laser is used as a seed for the next span, $N+1$. Then at Lab 2 or at the next repeater laser station, part of the light is reflected back by a Faraday mirror, and a beat note between the round-trip signal and the RL is detected on the photodiode PD3. After filtering, amplification and processing, the span $N+1$ can be stabilized by acting on a AOM for instance. That way, one can repeat step-by-step the ultra-stable laser.  

For sake of simplicity, lets assumes that the servo loop and the link compensation are ideal for the first fiber span, denoted $N$. One can derive that the phase of the remote laser signal on the Faraday mirror FM R1 is equal to the phase of the ultra-stable laser on FM 1. Considering the next fiber span denoted $N+1$, one finds similarly that the remote laser phase on the Faraday mirror FM 2 at the link output is equal to the phase of the remote laser on the Faraday mirror FM R2. Finally, the end-to-end phase difference $\Delta\phi$ is given by\,:
\begin{equation}
\Delta\phi(t)=\gamma\omega\cdot\left(\int_{\textrm{A}}^{\textrm{FMR1}}\Delta T(z,t)\,\textrm{d}z-\int_{\textrm{A}}^{\textrm{FMR2}}\Delta T(z,t)\,\textrm{d}z\right)
\end{equation}
where A corresponds to the position of the output coupler OC R1, and given the assumption that the input phase is defined on the Faraday mirror FM 1.  For an homogeneous temperature inside the interferometer, the response of the interferometer to a temperature step will be minimized when the equivalent fiber lengths between the local laser and the two Faraday mirrors of the station are made equal, {\it i.e.} $L_{\textrm{R1}}=L_{\textrm{R2}}$. 

We have checked experimentally this equation by exciting such an interferometer with a temperature step of 4\,K, and with the same experimental apparatus as described above. The experimental realization of such a well-balanced interferometer setup is quite difficult because the fiber paths include an AOM and a coupler, as detailed in\,\cite{Lopez:2012}.We plot the results of the experiment in Fig. \ref{Fig:TempSensiInterfero}c. From a multiple linear regression on the data, we found a temperature sensitivity of +1\,fs/K, a sensitivity to the time derivative of the temperature of -6.2\,ps/(K/s), and a time error drift of $-1\cdot 10^{-4}$fs/s. By comparison with the design we used before and reported in\,\cite{Lopez:2012}, there is a 7-times improvement for the temperature sensitivity.

Finally, we tested the 1\,fs/K optimized interferometer of Fig.\,\ref{Fig:TempSensiInterfero}(c) on our  540-km long-haul fiber link set between Paris and Reims\,\cite{Lopez:2012}, in order to check the impact on the long-term stability of the link. We compare the stability of the transferred phase obtained with our first generation interferometer, that has a measured sensitivity of 7\,fs/K, and with the new interferometer. The experimental setup is otherwise exactly the same and is reported in\,\cite{Lopez:2012}. After adequate filtering and amplification, frequency data are recorded with a dead-time free counter\,\cite{KplusK}, operated in $\Lambda$-mode and with a 1-s gate time. The time error and the corresponding MDEV are plotted in Fig.\,\ref{Fig:fig3-results540km}. The time error plot shows a 3-times narrower rms-trace for the new interferometer at short timescale (1-10\,s). The phase drift that we observed previously is also reduced, from $2\cdot 10^{-4}$fs/s for the 7\,fs/K interferometer, to $2.5\cdot 10^{-5}$fs/s for the low sensitivity interferometer. This is a reduction by a factor 8. Phase oscillations due to the air conditioning system are also much smaller and their impact on the stability plot is significantly reduced. Note that the oscillation frequencies are different, as the data were not taken at the same epoch of the year. With the low sensitivity interferometer, the MDEV reaches a floor at $3 \cdot10^{-20}$ after 10\,000\,s of integration time, which is 10-times better than with the previous setup. We also observe a factor 3 improvement on the short-term relative frequency stability. It could be an indication that the design of the interferometer matters at short time and leads to a better rejection of fast frequency fluctuations. However, the short-term stability strongly depends on the fiber phase noise PSD, which can vary significantly along time. Therefore further investigations are needed to corroborate this result. 

\section{CONCLUSION}

In this article, we have presented investigations on noise processes occurring in bi-directional optical fiber links. We described three setups, for frequency comparison, Sagnac interferometry, and frequency transfer respectively, and introduce a novel setup for two-way frequency comparisons in real-time. For all these setups, we addressed both the unsuppressed delay noise and the interferometric noise for fiber-based interferometers. We derived the unsuppressed delay noise for each setup in the case of spatial correlation that is not an impulse function.

We perform theoretical and experimental investigations on the interferometric noise, for which we introduced a linearization of the thermal induced fibre noise. This work provides handful analysis for designing low noise fiber-based interferometers for the three kinds of interferometers we addressed. We have shown that a well designed Sagnac interferometer, which is intrinsically symmetric, is therefore the easiest to realize and the less sensitive to temperature fluctuations in practice. We reported outstanding temperature sensitivity performance of .03\,fs/K, which is the lowest experimental value reported so far to the best of our knowledge. We reported the temperature sensitivity of a two-way setup to be 2.2\,fs/K. 

We realized a proof-of-principle experiment for real-time two-way frequency comparison and show excellent agreement of the experimental results with theory. We demonstrate finally an interferometer for cascaded fiber links with sensitivity as low as 1\,fs/K. We compare the phase measurement recorded with our previous interferometer and with the improved one. We showed that the noise floor of a 540-km link is as low as $3\cdot10^{-20}$ at 4\,000 s of integration time, which is the best relative instability value reported so far to our knowledge for an actively noise cancelled link, over such distance of many hundreds of km and in such short integration time. 

\appendix
\section{Detailed Calculation on Power Spectrum Density of Noise with Correlation Lengths\label{AppendixA}}
\subsection{Definitions}
The phase noise accumulated by an optical carrier propagating along a fiber link from an end $1$ at coordinate $0$ to an end $2$ at coordinate $L$ and measured in $2$ at time $t$ is\,:
\begin{equation}
\phi _{12} (t)= \int^L_0 \delta \varphi (z, t-T+z/v) dz
\end{equation}

where $z$ is the position coordinate along the fiber, $T=L/v$ is the propagation delay from 0 to $L$, $v=c/n$ is the speed of light in the fiber and $\delta \varphi (z,t)$ is the fiber noise per unit of length at position $z$ and time $t$.
Similarly the phase noise accumulated by the optical carrier propagating in the backward direction from the position $2$ to the position $1$ and measured at position $1$ at time $t$ is\,:
\begin{equation}\label{Eq:AccPhaseNoise}
\phi _{21} (t)= \int^L_0 \delta \varphi (z, t-z/v) dz
\end{equation}

The general approach followed here consists on calculating the phase noise PSD $S_{x}$ of a random signal by calculating the Fourier transform of the autocorrelation function of its phase noise $R_{x}(\tau)=\overline{x(t)x(t-\tau)}$, where the bar denotes a time average\,:
\begin{equation}
S_{x}(f) = \mathcal{F} (R_{x}(\tau) ) 
\end{equation}

\subsection{Free Fiber Noise PSD}
We consider the autocorrelation function $R_{12}(\tau)$ of $\phi_{12}(t)$\,:
\begin{equation}\label{Eq:R12}
\begin{split}
& R_{12}(\tau)= \overline{\phi_{12} (t)\cdot \phi_{12}(t-\tau)} = \\
& \iint^L_0 
\overline{\delta \varphi \left( z,t-T+\frac{z}{v} \right) \delta \varphi \left( z',t- \tau-T+\frac{z'}{v} \right)}
 dz dz'
\end{split}
\end{equation}

The correlation function $R$ inside the double integral can be expressed as a function of the difference of the linear coordinates  and of the relative delay\,:
\begin{equation}
\label{Eq:R}
R_{12}(\tau)=\iint_0^L R \left( z-z', \tau+\frac{z-z'}{v} \right) dz dz'
\end{equation}

We assume here that the space and time dependencies can be separated. We split $R$ into two different functions $R_t$ and $R_s$. 
\begin{equation}\label{Eq:Rsplitted}
R \left( z-z', \tau+\frac{z-z'}{v} \right) = R_t \left( \tau+\frac{z-z'}{v} \right) \cdot R_s(z-z') 
\end{equation}

We write $R_s$ as a gaussian with a deviation $\Delta L$ \,: 

\begin{equation}\label{Eq:R_s}
R_s(z-z')= e^{- \frac{1}{2} \left( \frac{z-z'}{\Delta L} \right) ^2 }
\end{equation}

We deduce the phase noise PSD at the end $2$ as\,:
\begin{equation}\label{Eq:S12_int}
\begin{split}
S_{12}(f) = \mathcal{F}( R_t(\tau)) \iint_0^L  e^ {  j 2 \pi f \cdot \frac{z-z'}{v} } e^ { -\frac{1}{2} \left( \frac{z-z'}{\Delta L} \right) ^2 } dz dz'
\end{split}
\end{equation}
We have  $S_{\delta \varphi}(f)=\mathcal{F}(R_t(\tau))$, which is the PSD of the phase noise per unit of length $\delta \varphi (z,t)$. Note that starting the calculations from the phase noise $\phi_{21}(t)$ leads exactly to the same result, so $S_{12}(f)=S_{21}(f)$.

We can approximate the limits of the inner integral of Eq. \ref{Eq:S12_int} with $\pm \infty$ on the assumption that $\Delta L << L$ and by neglecting the boundary effects. 
We set $x=z-z'$ and one obtains\,:
\begin{equation}
 \label{Eq:Inf}
 \begin{split}
  S_{12}(f) =S_{\delta \varphi}(f) \int_0^L \int_{-\infty}^{+ \infty} \exp j\left(2 \pi f \frac{x}{v}\right) e^{- \frac{1}{2} \left( \frac{x}{\Delta L} \right) ^2 } dx dz'&
 \end{split}
\end{equation}

We obtain finally\,:
\begin{equation}
S_{12}(f)= \Delta L L \sqrt{2 \pi} \cdot   e^ {  - \left( \frac{ \sqrt{2} \pi f \Delta L  }{v} \right)^2 }  S_{\delta \varphi}(f)
\end{equation}

\subsection{Two-Way Noise PSD}

The two-way phase noise at time $t$ is given by $\phi_{\textrm{tw}} (t)= \frac{1}{2} \left( {\phi _{12} (t) -\phi _{21} (t)}\right)$. From Eq. \ref{Eq:AccPhaseNoise} one has\,:
\begin{equation}
\phi_{\textrm{tw}} (t) =\frac{1}{2} \int^L_0 \left( \delta \varphi (z, t-T+\frac{z}{v}) - \delta \varphi (z, t-\frac{z}{v})  \right) dz
\end{equation}

A first order (and even a second order) Taylor expansion gives\,:
\begin{equation}
\label{Eq:TwoWayStandard}
\phi_{\textrm{tw}} (t)= \frac{1}{2} \int^L_0 \left( \frac{2z}{v}-T\right) \cdot \delta \varphi ' (z,t) dz
\end{equation}
where $\delta \varphi ' (z,t)$ is the time derivative of $\delta \varphi(z,t)$.

We calculate the autocorrelation function $R_{\textrm{tw}} (t)$ of the two-way fiber noise as\,:
\begin{equation}\label{Eq:R_TW}
\begin{split}
& R_{\textrm{tw}}(\tau)= \overline{\phi_{\textrm{tw}} (t)\cdot \phi_{\textrm{tw}}(t-\tau)} = \\
& \frac{1}{4} \iint^L_0 \left(\frac{2z}{v}-T\right)\left(\frac{2z'}{v}-T\right) 
\overline{\delta \varphi ' (z,t) \delta \varphi ' (z',t- \tau)}
 dz dz'
\end{split}
\end{equation}

Considering that\,\cite{Bendat:2012}\,: 
\begin{equation}
\label{Eq:Rt''}
\overline{\delta \varphi ' (z,t) \delta \varphi ' (z',t- \tau)} = - \frac{d^2}{d \tau^2} \overline{\delta \varphi  (z,t) \delta \varphi  (z',t- \tau)}
\end{equation}
we can use once again the  $R$ function introduced in Eq.\,\ref{Eq:Rsplitted} with $R(z-z',\tau)= \overline{\delta \varphi  (z,t) \delta \varphi  (z',t- \tau)} $. We split the $R$ function into one spatial-dependent term and one time-dependent term, $R(\tau)=R_t(\tau)  \cdot R_s (z-z') $. We describe again the spatial dependency with a gaussian function written as Eq.\,\ref{Eq:R_s}. 

One obtains\,:
\begin{eqnarray}
\label{Eq:R_TW_Split}
& R_{tw}(\tau)= &- \frac{1}{4} \iint^L_0 \left( 2z/v-T \right) \left( 2z'/v-T \right)\times\\
& & R_t''(\tau) e^{ - \frac{1}{2}\left( \frac{z-z'}{\Delta L} \right)^2} dz dz'\nonumber
\end{eqnarray}

where $R_t''(\tau)$ denotes the second derivative of $R_t(\tau)$. After some mathematics one obtains\,:
\begin{equation}
\label{Eq: -+inf}
\begin{split}
& R_{\textrm{tw}} (\tau) = -\frac{ {R_t''(\tau)}}{4}
\int_0^L \left( 2 z'/v-T \right)^2 \int^{+\infty}_{-\infty}  
e^{-\frac{1}{2} \left(\frac{x}{\Delta L } \right)^2} dx dz'
\end{split}
\end{equation}

which gives finally\,:
\begin{equation}
\label{Eq:Rtw_final}
\begin{split}
R_{\textrm{tw}}(\tau)=-R''_t(\tau) \sqrt{2 \pi} \Delta L v \frac{T^3}{12}
\end{split}
\end{equation}

The PSD of the two-way fiber noise is calculated as the Fourier transform of $R_{tw}(\tau)$\,:
\begin{equation}
\begin{split}
S_{\textrm{tw}}(f)= \frac{(2 \pi f T)^2}{12} \mathcal{F}(R_t(\tau)) \sqrt{2 \pi} \Delta L L
\end{split}
\end{equation}

The two-way PSD can be expressed as a function of the one-way PSD\,:
\begin{equation}
\begin{split}
&S_{\textrm{tw}}(f)=\frac{\left( 2 \pi f T \right)^2 }{12} e^{\left( \frac{\sqrt{2} \pi f \Delta L }{v} \right)^2 } S_{12}(f)
\end{split}
\end{equation}

\subsection{Actively Noise Compensated PSD}

We can approximate at first order the compensated fiber noise\,\cite{Williams:2008} as\,:
\begin{equation}
\phi_{\textrm{anc}}(t)= \int_0^L \frac{z}{v} \delta \varphi ' (z,t) dz
\end{equation}

The corresponding auto-correlation function is\,:
\begin{equation}
\begin{split}
& R_{\textrm{anc}}(\tau)= \iint_0^L \frac{z}{v}\frac{z'}{v} \overline{\delta \varphi ' (z,t) \delta \varphi ' (z',t)}dz dz'
\end{split}
\end{equation}

Following the same steps of Eq.\,\ref{Eq:Rsplitted}, \ref{Eq:Rt''}-\ref{Eq:Rtw_final} we obtain\,:
\begin{equation}
\label{Eq:Ra}
R_{\textrm{anc}}(\tau)=- R''(\tau) \int_0^L \left( \frac{z'}{v}\right) ^2 \int_{-\infty}^{+\infty} e^{-\frac{1}{2} \left( \frac{x}{\Delta L} \right)^2  }dx dz'
\end{equation}
which leads to the expression\,:
\begin{equation}
R_{\textrm{anc}}(\tau)= -R''_t(\tau) \sqrt{2 \pi} \Delta L v \frac{T^3}{3}
\end{equation}
which is four times the expression of $R_{\textrm{tw}}(\tau)$ given by Eq.\,\ref{Eq:Rtw_final}. 

The noise PSD of the ANC setup can be expressed as a function of the one way free fiber noise PSD $S_{12}(f)$\,:
\begin{equation}
S_{\textrm{anc}}(f)=\frac{\left( 2 \pi f T \right)^2 }{3} e^{\left( \frac{\sqrt{2} \pi f \Delta L }{v} \right)^2 } S_{12}(f)
\end{equation}

\section{Mixed Two-way and ANC noise rejection\label{AppendixB}}
\subsection{Two-way + Active Noise Compensation}

We are now calculating the effects on the fiber noise of a composite two-way + Actively Noise Compensated  setup. 

We consider first a two-way scheme combined with one single active noise compensation set at one end, that we choose to be at end $1$ for instance. The phase noise accumulated by the optical carrier propagating along the link in the forward direction from the end $1$ and exiting the fiber at output end $2$ at time $t$ is\,:
\begin{equation}
\phi _L(t)=\int_0^L \delta \varphi(z,t-T+z/v) dz + \phi_{c}(t-T)
\end{equation}
where\,\cite{Bercy:2014a}\,:
\begin{equation}\label{Eq:phi_corr}
\phi_c (t-T)=-\frac{1}{2}\int_{0}^{L} \left( \delta\varphi(z,t-\frac{z}{v})+\delta\varphi(z,t-2T+\frac{z}{v})\right) dz
\end{equation}

The phase noise accumulated by the optical carrier propagating in the backward direction from the end $2$ and exiting the fiber at the end $1$ at time $t$ is\,:
\begin{equation}
\phi _0 (t)=\int_0^L \delta \varphi(z,t-z/v) dz + \phi_{c}(t)
\end{equation}

Thus the two-way phase noise with compensation $\phi_{tw+c}(t)=1/2(\phi _L (t)-\phi _0 (t))$ at time $t$ is given by\,:
\begin{equation}
\phi _{tw+c}(t)=\int_0^L \frac{z}{v}\delta \varphi '(z,t) dz
\end{equation}

Assuming a negligible correlation length for the sake of simplicity, the phase noise PSD at the output end $2$ is again calculated as the Fourier transform of the
$\phi_{tw+c}(t)$ auto-correlation function, and one finds\,: 
\begin{equation}
S_{tw+c}(f)=\frac{(2\pi f T)^{2}}{3}S_{12}(f)
\end{equation}
The power spectral density of noise is therefore four times more than that of a post-processed two-way without active compensation. 

\subsection{Two-way + Double Noise Compensation}

The second approach is as follow : in a first step the fiber noise is removed by post-processing at both ends of the fiber, and in a second step the two resulting phases are subtracted with a two-way technique.
We can write the phases recorded at the positions $1$ and $2$ (at coordinates $0$ and $L$ respectively) at time $t$ as\,\cite{Williams:2008}\,:
\begin{equation}
\begin{split}
&\phi _O(t)=\int_0^L \delta \varphi(z,t-z/v)dz + \phi_{c,2}(t-T)\\
&\phi _L(t)=\int_0^L \delta \varphi(z,t-T+z/v)dz + \phi_{c,1}(t-T)
\end{split}\end{equation}

where the terms $\phi_{c,2}$ and $\phi_{c,1}$ are the two fiber noise corrections calculated in $L$ and $0$ and have similar expression as the one given by Eq\,\ref{Eq:phi_corr}. The corrections terms are applied in post processing to the signals registered at the opposite end of the link.
Thus the two-way phase noise with double compensation $\phi_{tw+2c}(t)=1/2(\phi _L (t)-\phi _0 (t))$ at time $t$, is\,:
\begin{equation}
\begin{split}
\phi_{tw+2c}(t)=&\frac{1}{2} \left[ \phi_0(t)-\phi_L(t) \right]=\phi_{\textrm{tw}}(t)+ \\
+&\frac{1}{2} \left[ \phi_{c,2}(t-T)-\phi_{c,1}(t-T) \right]
\end{split} 
\end{equation}

The first term derives from the subtraction of the two free fiber noises integrated from $0$ to $L$ and from $L$ to $0$ and corresponds to the standard two-way signal of Eq \ref{Eq:TwoWayStandard}. The second term derives from the interaction of the two ANC corrections and can be developed as follows\,:

\begin{equation}
\begin{split}
&\frac{1}{2} \left[ \phi_{c,2}(t-T)-\phi_{c,1}(t-T) \right]=\\
 = &\frac{1}{2}\int_0^L \left[ \delta \varphi(z,t-z/v)+\delta\varphi(z,t-2T+z/v)\right] dz+\\
-&\frac{1}{2}\int_0^L \left[ \delta \varphi (z,t-T+z/v)+\delta\varphi(z,t-T-z/v)\right] dz
\end{split}\end{equation}

We can easily calculate that this term is zero with a first order development. We conclude that implementing an additional double ANC-like compensation to a two-way link does not improve the overall noise rejection.

\subsection{Two-way Real-Time PSD}

In a real-time two-way implementation, the one way fiber noise measured at one end is corrected with half of the round trip fiber noise measured at the same end of the link. For example, the round trip fiber noise measured at the end $1$ (coordinate $0$) and time $t$ is\,:

\begin{equation}
\label{round trip noise}
\phi_{212}(t)=\int_0^L \delta \varphi(z,t-z/v)dz+\int_0^L \delta \varphi (z,t-2T+z/v)dz
\end{equation}

So, the real-time two-way signal is, with a first order approximation\,:
\begin{equation}
\phi_{\textrm{rt-tw}}(t)=\phi_{21}(t)-\frac{1}{2}\phi_{212}(t)\approx\int_0^L \frac{L-z}{z}\delta \varphi '(z,t)dz
\end{equation}

We can proceed to calculate the autocorrelation of $\phi_{\textrm{real}}(t)$ with the method described in the previous paragraphs\,:
\begin{equation}
R_{\textrm{rt-tw}}(\tau)=\iint_0^L \frac{L-z}{v}\frac{L-z'}{v} \overline{ \delta \varphi'(z,t)\delta \varphi'(z',t-\tau)}dz dz'
\end{equation}

Following the same steps from Eq.\,\ref{Eq:Rsplitted}, \ref{Eq:Rt''}-\ref{Eq:Rtw_final} we obtain the following result\,:

\begin{equation}
R_{\textrm{rt-tw}}(\tau)= -R''_t(\tau) \sqrt{2 \pi} \Delta L v \frac{T^3}{3}
\end{equation}

that is exactly the same obtained in Eq \ref{Eq:Ra}. We can conclude that the real-time two-way noise compensation system leads to the same reduction on the free fiber noise of the ANC frequency dissemination setup\,:

\begin{equation}
\begin{split}
&S_{\textrm{rt-tw}}(f)=\frac{\left( 2 \pi f T \right)^2 }{3} e^{\left( \frac{\sqrt{2} \pi f \Delta L }{v} \right)^2 } S_{12}(f)
\end{split}
\end{equation}

\begin{acknowledgments}
The authors thank Emilie Camisard, Thierry Bono and Patrick Donath at the GIP Renater,  for giving the opportunity to use the RENATER network.

This work is supported by the European Metrology Research Programme (EMRP) under SIB-02 NEAT-FT. The EMRP is jointly funded by the EMRP participating countries within EURAMET and the European Union. This work is supported by the Labex FIRST-TF, the french spatial agency CNES, IFRAF-Conseil R\'{e}gional Ile-de-France and Agence Nationale de la Recherche (ANR BLANC 2011-BS04-009-01 and REFIMEVE+).
\end{acknowledgments}

\bibliographystyle{unsrt}
\bibliography{Tackling}

\end{document}